**MEEPTOOLS: A maximum expected error based FASTQ read filtering and trimming toolkit**

Vishal N. Koparde[1*] (vnkoparde@vcu.edu)

Hardik I. Parikh[2] (parikhhi@vcu.edu)

Steven P. Bradley[2] (bradleysp@vcu.edu)

Nihar U. Sheth[1] (nsheth@vcu.edu)

[*]Corresponding Author

[1]Center for the Study of Biological Complexity, Virginia Commonwealth University, Richmond, Virginia, USA

[2]Department of Microbiology and Immunology, Virginia Commonwealth University, Richmond, Virginia, USA

**Availability and implementation**: MEEPTOOLS is open source and available at https://github.com/nisheth/meeptools.

**Abstract**

Next generation sequencing technology rapidly produces massive volume of data and quality control of this sequencing data is essential to any genomic analysis. Here we present MEEPTOOLS, which is a collection of open-source tools based on maximum expected error as a percentage of read length (MEEP score) to filter, trim, truncate and assess next generation DNA sequencing data in FASTQ file format. MEEPTOOLS provides a non-traditional approach towards read filtering/trimming based on maximum error probabilities of the bases in the read on a non-logarithmic scale. This method simultaneously retains more reliable bases and removes more unreliable bases than the traditional quality filtering strategies.

**Introduction**

Interpreting DNA sequences is a vital component of all biological research efforts(Rizzo and Buck, 2012)(Su et al., 2011)(Xuan et al., 2013). High throughput DNA sequencers or next-generation sequencing (NGS) platforms like Illumina have enabled researchers to generate sequence data at a rate, which outpaces Moore's law and is approximately doubling annually(Gullapalli et al., 2012). NGS data has been utilized in a broad range of applications, like *de novo* assemblies of genomes and transcripts(Ekblom and Wolf, 2014); mapping of reads to existing reference genomes to determine variations (SNPs, etc.)(Kimura and Koike, 2015)(Koboldt et al., 2012) and to explore expression levels (RNA-Seq)(Wang et al., 2009); haplotype inference(Aguiar and Istrail, 2012); detecting DNA methylation(Peters et al., 2015); etc. The accuracy of these techniques is a sensitive function of the quality of the NGS data used as input. Hence, filtering and/or trimming of NGS data, which is a widely accepted and practiced non-trivial preprocessing step for all NGS data analysis, is critical and may have a significant effect on the scientific results.

Illumina reads, which are encoded in the FASTQ file format, augment an ASCII quality score to each base called by the sequencer. This quality score or PHRED $Q$(Ewing and Green, 1998) is a way of encoding the probability $p$ that the corresponding base call is incorrect.

$$p = 10^{-Q/10}$$

or

$$Q = -10 \, log_{10} \, p$$

Typically, $Q$ ranges from 0 to 41, corresponding to a 100% to a 0.007943% chance of an incorrect base call. Most read filtering and trimming tools, like FASTX-Toolkit, NGS-QC Toolkit(Patel and Jain, 2012), PRINSEQ(Schmieder and Edwards, 2011), Trimmomatic(Bolger et al., 2014), etc., assess the overall quality of a read by calculating the average $Q$ value across all bases in the read. This, we believe, is not a good indicator of overall read quality as a $Q40$ base call is not twice as good as a $Q20$, but 100 times less likely to be incorrect as the PHRED $Q$ is logarithmically related to the error probability of base-calling. Hence, we introduce MEEPTOOLS, which is capable of filtering and trimming FASTQ reads based on maximum expected error in base calling, an approach similar to one adopted by USEARCH(Edgar and Flyvbjerg, 2015).

**MEEPTOOLS software**

MEEPTOOLS is an open-source software written in the C-programming language to filter, trim and sort single and paired-end FASTQ read datasets. It can also be used to simply access the quality of the reads by generating relevant statistics. Maximum expected error (MEE) in a read is the sum of all the expected error probabilities of all the bases in that read, and when MEE is expressed as a percentage of read length, it denotes the maximum number of probable incorrect base calls per every 100 bases in the read. Hence, we propose the MEEP score:

$$MEEP = 100X \sum_{i=1}^{rl} p_i \bigg/ rl$$

where $rl$ is the read length, $p_i$ is the error probability at position $i$.

MEEPTOOLS has the following subprograms:

a. **stats**: This generates MEEP related statistics along with some basic statistics about the FASTQ file like number of reads, number of bases, minimum/maximum/average read length, average read $Q$, overall MEEP for the entire file, number of reads with MEEP less than 1 or 2 (MEEP1/MEEP2), number of reads with average read $Q$ greater than $Q_{threshold}$, etc. We propose that metrics like overall MEEP and percentage of the dataset satisfying the MEEP1 criterion give shed more light on the overall quality of the dataset and can quantify the comparison of two different datasets.

b. **append**: FASTQ file format has a comment section for each read. This subprogram allows user to append MEE, MEEP and average read $Q$ information to this comment section, making it easily available for downstream analysis software.

c. **filter**: A filtered subset of reads, where all the reads have MEEP score less than a user specified threshold, can be generated using this subprogram. A filtering based on read-length can also be concurrently performed.

d. **trim**: Most of the commonly used trimming tools are either based on a) running sum algorithms or b) window base algorithms. MEEPTOOLS considers trimming to be a "minimum subarray problem" and implements a modified version of Kadane's algorithm(Bentley, 1984) to efficiently find the "sub-read", which meets the user defined

MEEP and readlength cutoffs. If readlength, $rl$, does not meet the thresholds, then two sub-reads of readlengths $rl - 1$ are considered, and if even those do not meet the thresholds, then three subreads of readlengths $rl - 2$ are considered. This iterative process terminates with either finding the sub-read, which meets the thresholds, or moving on to the next read as readlength becomes smaller than the threshold. As we are searching for a "sub-read", the actual trimming could occur from 3' or 5' or both ends of the original read.

  e. **sort**: The sort subprogram reads in the FASTQ file and generates a new FASTQ file where the reads are sorted in ascending order of their respective MEEP scores. Thus, the read with the lowest expected errors is at the top of the output file and the read with the most expected errors is at the bottom.

MEEPTOOLS has an option to truncate the output from all subprograms after a certain number of reads; thereby generating a subset of reads meeting a predefined criteria. Thus, we can generate a subset of million highest quality reads by truncating the output from *sort* subprogram to a million reads.

**Conclusions**

The main motivation towards development of MEEPTOOLS is to achieve comprehensive quality control of NGS data via filtering and trimming of reads based on error probabilities in non-logarithmic space, *in lieu of* PHRED $Q$. MEEPTOOLS strategically retains high quality sub-read from a low average quality read, thereby resulting in higher read survival rate when compared with traditional trimming techniques. This systematic elimination of lower quality bases will no doubt bolster the accuracy of any downstream analysis. Hence, we believe that MEEPTOOLS performs an effective and efficient job at read trimming and filtering, and can replace traditional approaches in existing NGS data processing and analysis pipelines. Our tests indicate that MEEPTOOLS can process 183-470,000 reads per minute while trimming and 207-770,000 reads per minute while filtering based on the dataset read lengths. See Supp. Tables 1-3 for details.

**Acknowledgements**

HP and SB are supported by the NIH Common Fund Human Microbiome Project (HMP) program through grant 8U54HD080784 to G Buck, J Strauss, and K Jefferson. We gratefully acknowledge the technical and philosophical discussions with our colleagues Myrna Serrano, Gregory Buck, and Shaun Norris. We are thankful to Jonathan Kindberg for introducing and helping our group with tracking of agile project management process, which enabled us to produce this and other publications in quick time.

*Conflict of interest:* None.